\documentclass[aps,twocolumn]{revtex4}
\usepackage{epsfig,graphics,amssymb,amsmath}

\def\d{{\rm d}}

\def\deg{$^\circ$}

\def\be{\begin{equation}}
\def\ee{\end{equation}}

\def\besub{\begin{subeqnarray}}
\def\eesub{\end{subeqnarray}}

\begin{document}

\title{\textbf{Evaporation-driven assembly of colloidal particles}}
\author{Eric Lauga \& Michael P. Brenner}
\affiliation{Division of Engineering and Applied Sciences,
Harvard University,\\ 29 Oxford Street, Cambridge, MA 02138.}
\date{\today}

\begin{abstract}
Colloidal particles absorbed at the interface of a liquid droplet
arrange into unique packings during slow evaporation (Manoharan
{\it et al.} {\it Science} {\bf 301} 483-487). We present a
numerical and theoretical analysis of the packing selection
problem. The selection of a unique packing arises almost entirely
from geometrical constraints during the drying.
\end{abstract}

\maketitle

Manoharan {\it et al.} \cite{Vinny} recently presented an
ingenuous method for fabricating clusters of small particles into
precise configurations. Polystyrene spheres (diameter 844 nm) were
dispersed in a toluene-water emulsion with each oil droplet
containing a low number $N$ of spheres.  The toluene was then
preferentially evaporated, forcing the particles to come together
into compact clusters. Surprisingly, the final particle packings
were unique: the observed packings for $N\leq 11$ closely
correspond to those previously identified \cite{Sloane} as
minimizing the second moment of the particle distribution, ${\cal
M}=\sum_i||{\bf r}_i-{\bf r}_0||^2$, where ${\bf r}_0$ is the
center of mass of the cluster.

The fact that such a simple process leads to precision assembly at
the submicron scale points to exciting possibilities for
controlling the assembly of more general objects \cite{review}.
The goal of this Letter is to understand the physical principles
underlying the observations of Manoharan {\it et al.}
\cite{Vinny}. Why are the final packings unique? (Why) do they
minimize the second moment? What physical parameters do these
results depend upon? We first present numerical simulations of
hard spheres on an evaporating liquid droplet for a wide range of
liquid-solid contact angles: for each contact angle the
simulations reproduce the final packings of \cite{Vinny}. We then
demonstrate that the uniqueness of the packings, as well as their
connection to minimal moment structures, can be understood from
purely geometrical considerations. The arguments suggest a
methodology for creating new packings, which we confirm through
numerical simulations.

\begin{figure}[b]
\centering
\includegraphics[width=.45\textwidth]{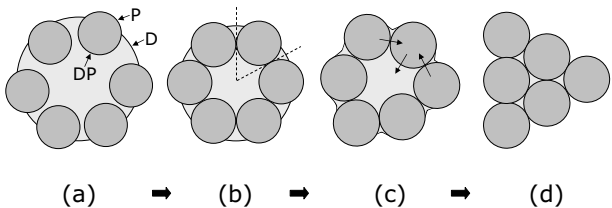}
\caption{Schematic two-dimensional representation of the drying
experiment: (a) particle configuration when the droplet volume is
above the critical volume; (b) critical packing; (c) rearrangement
below the critical packing with both capillary and contact forces
acting on each particle; (d) final packing.} \label{schematic}
\end{figure}

{\it Numerical Simulations.}  The experiments suggest the
following theoretical problem: for a given liquid volume, the
particle configuration is determined from minimizing the total
surface energy
\begin{equation}\label{surf}
U_\Sigma=\gamma_D \int_{D} \d S+ \gamma_{DP} \int_{DP} \d S+
\gamma_P \int_{P}\d S,
\end{equation}
while respecting excluded volume constraints between the
particles. Here $D$, $DP$ and $P$ refer to the droplet surface,
droplet-particle interface and particle surface respectively (see
Figure \ref{schematic}).

\begin{figure}[b]
\centering
\includegraphics[width=.48\textwidth]{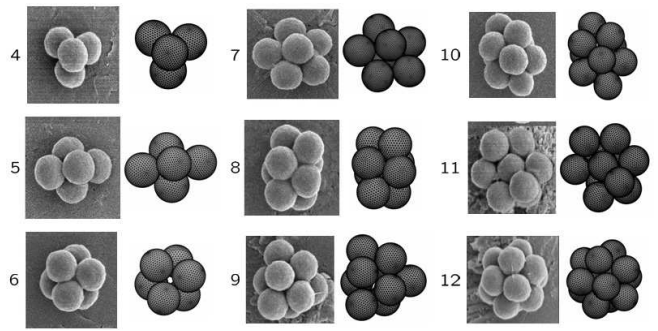}
\caption{Comparison of the experimentally observed packings (left)
with those obtained by numerical simulations (right) as a function
of the number of spherical particles.} \label{comparison}
\end{figure}

Numerical simulations of this problem are performed using Brakke's
Surface Evolver \cite{Brakke}, a program which determines the
equilibrium configuration of deformable surfaces given the
definition of an energy. The colloidal spheres are modelled as
liquid droplets with high surface tension, typically 1 to 2 orders
of magnitude larger than the main droplet, in order to penalize
non-spherical deformations of their shape. Interfacial tension
between the droplet and the particles are chosen appropriately in
order to satisfy Young's law at the solid-liquid contact line,
$\gamma_P = \gamma_D \cos\theta + \gamma_{DP}$, where $\theta$ is
the equilibrium contact angle. Non-interpenetrability is enforced
with an excluded volume repulsion energy $U_R$ acting between the
centers of the spheres \footnote{We have used potentials ranging
from exponential $U_R\sim a\exp p (1-d/d_0)$, to power law,
$U_R\sim a (d_0/d)^p$, where $d$ ($d_0$) is the distance between
two particles (diameter of a particle). The results are
insensitive to the exact functional form of $U_R$.}; $U_R$
dominates when at least two spheres overlap by one percent.

The particles are initially positioned randomly on the droplet.
The droplet volume is then slightly decreased (by one percent or
less) and the particles rearrange to a new equilibrium, minimum of
${\cal U}=U_\Sigma+U_R$. This procedure replaces the evaporation
dynamics by a series of equilibrium problems and therefore mimics
the low evaporation rate limit of the experiments.

We find that the packings obtained numerically are unique and
agree with those obtained by Manoharan {\it et al.} \cite{Vinny},
over the range of contact angles tested ($10^{\circ} \le \theta
\le 170^{\circ}$) and initial conditions. A comparison of the
final computational and experimental packings is illustrated in
Figure \ref{comparison}. In all cases, the final second moment
obtained numerically differs by less than 0.5 \% from the minimum
moment packings of \cite{Sloane}. The agreement verifies that the
experimental packing problem corresponds to the energy
minimization posed above and that other particle-particle
interactions (electrostatic, van der Waals...) are not essential
to generate the final packings.

\begin{figure}[t]
\includegraphics[width=.48\textwidth]{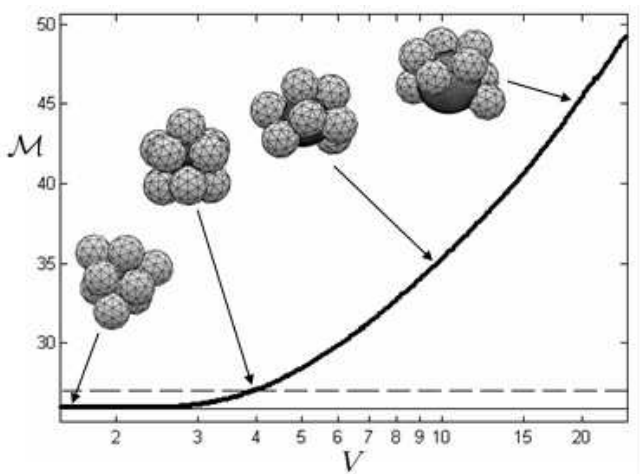}
\centering \caption{Numerical evolution of the second moment as a
function of the droplet volume for $N=9$ (both in units of the
particle radius). The final second moments are 25.899 (theory) and
25.946 (simulations). Insert:  packings observed during the drying
process when ${\cal M}=25.946, 27.016, 35.014,45.058$. Solid line:
minimal second moment; dashed line: second moment at the critical
packing (${\cal M}=27$).} \label{evolution}
\end{figure}

There is one noteworthy difference between the simulations and the
experiments. In the simulations, the cluster of spheres evolves
smoothly and continuously throughout the drying process (c.f. Fig.
\ref{evolution} for $9$ particles). In contrast, experiments
\cite{Vinny} showed a discontinuous transition to the final
packing, with an abrupt change in the second moment at a critical
volume. The experimental discontinuities must arise from
experimental features not present in the simulations, probably
contact angle hysteresis.

{\sl The Critical Volume.} We now study theoretically the packing
selection problem. For sufficiently large liquid volumes, the
minimum energy solution is a spherical droplet with noninteracting
force-free particles (Fig. \ref{schematic}a). There exists a
critical volume $V_c$ below which the droplet cannot remain
spherical (Fig. \ref{schematic}b); below this critical volume each
particle is acted upon by capillary forces. At the critical volume
$V_c$, there is a critical packing of particles, which can be
characterized as follows.

The interactions between particles on the surface of a sphere are
equivalent to the steric interactions between the ``cone of
influence'' of each particle, defined as the cone originating from
the droplet center and tangent to the particle (Fig.
\ref{schematic}b).  These are also equivalent to the interactions
between the intersection of these cones with the droplet, which
are circles. Consequently, packing spherical particles at the
critical volume is equivalent to packing circles on a sphere, a
mathematical problem with rich history
\cite{Melnyk,Clare,Kottwitz,Sloane_circles} and for which
numerical solutions have been proposed up to $N={\cal O}(100)$.
Usually, the circle packings for a given $N$ are unique, with two
types of exceptions. For $N=5,19,20,23,26,28,29\dots$ the circle
packing has continuous degrees of freedom (where at least one
circle is free to ``rattle'') \cite{Kottwitz}; for $N=15$ two
different configurations (15a and 15b) lead to the densest
packings with equal surface density \cite{Kottwitz}.

Hence, provided that there are no kinetic traps, identical
particles will arrange themselves into the circle packing at the
critical volume. Fig. \ref{evolution} confirms this conclusion in
numerical simulations for $N=9$. Note that changing the contact
angle of a particle changes the size of its circle of influence
and the droplet radius  by the same amount, so the circle packing
problem is unchanged and the critical packing of particles at the
critical volume is independent of wetting characteristics; note
however that the critical droplet volume is contact angle
dependent.

\begin{table}[t]
\begin{tabular}{rcccclcccclccc}
$N$&$n_c$&$n_f$&$n_m$&\phantom{12345}&$N$&$n_c$&$n_f$&$n_m$&\phantom{12345}&$N$&$n_c$&$n_f$&$n_m$\\
\hline
{\bf 4}&6&1&1           &&{\bf 13} &24&1&1        &&{\bf 21}&40&1&1                      \\
{\bf 5}&6&1&3           &&{\bf 14} &28&3&1        &&{\bf 22}&42&1&1                      \\
{\bf 6}&12&3&1          &&{\bf 15a}&30&3&1        &&{\bf 23}&43&2&3                         \\
{\bf 7}&12&1&1          &&{\bf 15b}&30&3&1        &&{\bf 24}&60&15&1                        \\
{\bf 8}&16&3&1          &&{\bf 16} &32&3&1        &&{\bf 25}&48&1&1                         \\
{\bf 9}&18&3&1          &&{\bf 17} &34&3&1        &&{\bf 26}&46&1&5                       \\
{\bf 10}&19&2&1         &&{\bf 18} &34&1&1        &&{\bf 27}&52&1&1                       \\
{\bf 11}&25&6&1         &&{\bf 19} &34&1&3        &&{\bf 28}&52&1&3                      \\
{\bf 12} &30&9&1        &&{\bf 20}&39&6&5         &&{\bf 29}&54&1&3 \\
\end{tabular}
\caption{Characteristics of critical packings of spherical
colloidal particles as a function of their number $N$: number of
contacts ($n_c$), number of independent forces ($n_f$) and number
of admissible modes of rearrangement ($n_m$).}\label{table}
\end{table}

{\sl How do the particles rearrange when $V<V_c$?} The energy
minimization problem suggests that we must find the particle
configuration which minimizes $U_\Sigma$ under inter-penetrability
and contact angle constraints.  Owing to the complexity of solving
for the liquid surface of constant mean curvature, this at first
appears to be an extraordinarily difficult theoretical problem.
However, we have found that in fact the constraints associated
with packing of particles are sufficient to uniquely determine the
initial rearrangements of the particles.

Let us suppose the droplet volume is reduced by a small amount
$\delta V \ll V_c$. Deviations of the droplet interface from
spherical lead to capillary forces (${ \bf F}_i$) on each
particle. Since every particle must be in force equilibrium, these
forces must be balanced by contact forces (${\bf f}_{ji}$) between
the particles (Figure \ref{schematic}c):
\begin{equation}\label{equilibrium}
{\bf F}_i + \sum_{j\in{\cal C}(i)} {\bf f}_{ji}={\bf 0},
\end{equation}
where ${\cal C}(i)$ denotes the set of particles in contact with
sphere $i$.

Let us characterize the number of ways the particles can rearrange
to accommodate this change in volume. Each particle has three
degrees of freedom and the droplet has one (the value of its mean
curvature or pressure), so there are  $3N+1$ degrees of freedom.
The constraints are of three types: (a) solid body rotation does
not modify the packing (3 constraints); (b) the particles cannot
overlap, ($n_c$ constraints, where $n_c$ is the number of contacts
at the critical packing); (c) forces have to balance (Eqn.
\ref{equilibrium}).

Equation \eqref{equilibrium} implies nontrivial constraints.
Indeed, suppose we try to solve \eqref{equilibrium}  for the $n_c$
contact forces ${\bf f}_{ji}=f_{ji}{\bf e}_{ji}$ (${\bf
e}_{ji}={\bf e}_i-{\bf e}_j$, where ${\bf e}_i$ is the unit vector
directed from the droplet center to the center of the particle).
Equation \eqref{equilibrium} has $3N$ components, and $n_c\sim 2N$
unknowns (Table I). Consequently, solutions exist only if
compatibility relations are satisfied between the ${\bf F}_i$.
Since capillary forces depend on the position of the particles,
these equilibrium considerations constrain the rearrangement of
the particles. Geometrical constraints have also played an
important role in understanding force propagation in granular
packings \cite{Witten}.

\begin{table}[t]
\begin{tabular}{lllllllllll}
$N$ &${\cal M}_{\rm exp}$& ${\cal M}_m$ & ${\cal
M}_{2}$&\phantom{123456}& $N$ &${\cal M}_{\rm exp}$& ${\cal M}_m$
& ${\cal M}_{2}$\\
\hline
${\bf 3}$   &4&4&4                 &&   ${\bf 9}$ &25.899&25.899&25.899\\
${\bf 4}$   &6&6&6                 &&   ${\bf 10}$&31.828&31.828&31.828 \\
${\bf 5}$   &9.333&9.333&9.333     &&   ${\bf 11}$&37.835&37.929&37.835 \\
${\bf 6}$   &12&12&12              &&   ${\bf 12}$&43.416&43.416&42.816 \\
${\bf 7}$   &16.683&17.100&16.683  &&   ${\bf 13}$&51.316&52.690&47.701 \\
${\bf 8}$   &21.157&21.657&21.157  &&   ${\bf 14}$&59.225&60.279&54.878 \\
\end{tabular}
\caption{Final second moment of the model (${\cal M}_m$) compared
with the final second moment in the experiment \cite{Vinny} if the
particles are assumed to be perfectly spherical  (${\cal M}_{\rm
exp}$) and with the minimum second moment (${\cal
M}_{2}$).}\label{table2}
\end{table}
Close to the critical volume, the capillary forces are given by
${\bf F}_i={F}_i{\bf e}_i$.  From the $N$ scalar forces $F_i$,
equation \eqref{equilibrium} shows that only a subset $n_f$ can be
chosen independently, {\it i.e.} equilibrium of each particle lead
to $N-n_f$ additional constraints; $n_f$ is found by computing the
rank of the compatibility matrix in \eqref{equilibrium}
\footnote{Since $\sum_i{\bf F}_i={\bf 0}$, we get necessarily that
$n_f\leq N-3$.} .

The total number of admissible modes of rearrangement $n_m$ for
the colloidal particles is found by subtracting the number of
constraints from the number of degrees of freedom: we find
$n_m=2N-2+n_f-n_c$. For a given $N$, $n_m$ is entirely determined
by the geometry of the critical packing. The results are displayed
in Table \ref{table}. Whenever the circle packing is unique, we
find that $n_m=1$. When $n_m>1$, we find that the number of modes
is always correlated with the presence of rattlers. If a total
number of continuous degrees of freedom $n_d$ exist in the circle
packing ($n_d$=2 or 1 per rattler depending on if it completely
free, as in $N=19$ or constrained in a slot, as in $N=5$), we
always find that $n_m=1+n_d$. Since $n_d\neq 0$ indicates that
there exists $n_d$ force-free surface modes for the packing, we
obtain therefore that there exists a unique mode of rearrangement
for all particles which are non-rattlers. In experiments, the
degeneracy in the circle packing problem is chosen by additional
information: for example in \cite{Vinny}, because the particles
are charged only on the side exposed to the water, there is a weak
dipolar repulsive force between the particles which breaks the
degeneracy \cite{Pieranski} \footnote{The strength of the dipolar
interaction on $n_c$ contacts is about $U_{\rm elec}\sim n_c
p^2/\epsilon_r\epsilon_0 a^3$ with a dipole moment $p\sim \sigma
a^2/\kappa\sqrt{\epsilon_r}$ \cite{Hurd} where $a$ is the particle
size, $\epsilon_r$ the dielectric constant of water and $\sigma$
the surface charge of the colloids with zeta potential $\zeta$ and
screening length $\kappa^{-1}$, $\sigma \sim \zeta \epsilon_0
\epsilon_r \kappa$; therefore $U_{\rm elec}\sim n_c\zeta^2
\epsilon_0 a$. By comparison, the change in surface energies
$\delta U_\Sigma$ due to a translation of order $\delta a$ of the
contact line of each particle and the creation of menisci is
$\delta U_\Sigma\sim N \gamma_D a \delta a$. Since $n_c\sim 2N$,
and with the values $\zeta\approx 100$ mV, $a\approx 1\,\mu$m and
$\gamma_D \approx 10^{-2}$ N.m$^{-1}$, we get that $U_{\rm elec}
\geq U_\Sigma$ when $\delta a \leq 10^{-5} a \ll 1$ nm. Hence, as
soon as menisci appear, the dipolar repulsion appears only as a
secondary minimization problem, governing the location of
particles after surface energies are minimized.}.

This result implies that there is only a single set of  $\{F_i\}$
that is consistent with all the constraints. This mode is
independent of surface energies, and depends only on geometry.
Capillarity does enter into the problem in relating the force
$F_i$ to the displacement $\delta r_i$ of the $i^{th}$ sphere. If
radius of curvature of the droplet changes from $R$ to $R +\delta
R$ then volume conservation of the droplet implies that $\delta
r_i$ and $\delta R$ are related through $A \sum_i \delta r_i +
(4\pi R^2-N A) \delta R=0$, where $A$ is the wetted area of the
particles. The capillary force $F_i$ is then given by
\begin{equation}
{F}_i=-2 \pi \gamma_D \cos\beta \left({\delta r_i}+
\frac{A{\cos\beta}}{4 \pi R^2-NA}\sum_i \delta r_i
\right),\label{fd}
\end{equation}
where $\alpha$ is the dry angle of the particle on the critical
packing, $\theta$ the equilibrium contact angle, and
$\beta=\alpha-\theta$. This formula is asymptotically valid in the
limit $\delta V \to 0$ so that deviations from a spherical cap
droplet are small.

\begin{figure}[t]
\centering
\includegraphics[width=.45\textwidth]{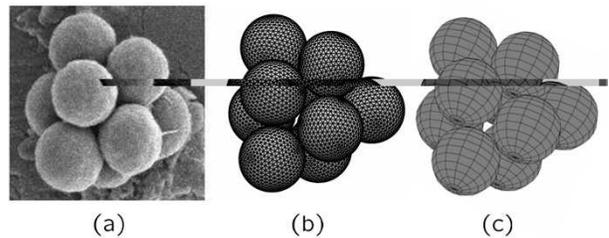}
\caption{Comparison for $N=9$ between the packing given by (a)
experiments, (b) simulations and (c) model.} \label{model}
\end{figure}

The above results apply just below $V_c$. However, the general
principle can be iteratively applied below the critical volume:
starting from the spherical packing, we decrease the droplet
volume by small increments and, assuming equation (\ref{fd})
continues to hold, we calculate the corresponding incremental
particle rearrangement consistent with all the constraints. At
each step in the iteration it is necessary to recompute the mode
$\{F_i\}$; for every packing there is a unique choice that is
consistent with the constraints. This process iterates until the
final equilibrium configuration is reached. During the process the
packing changes substantially from the initial disk packing--
typically multiple new contacts are added.

The results of the model are displayed in Table \ref{table2} and
in Figure \ref{model}.  The model reproduces accurately the final
experimental packings for $N\leq 6$ and $9\leq N \leq 14$; in
particular, the non-convex packing for $N=11$ is well predicted by
the model. There are small differences in the final packings for
$N=7$ and $8$, likely arising from deviations of the capillary
force-particle displacement relationship from the linear law
(\ref{fd}). Overall, the model leads to a dimensionless error in
second moment with the experimental packings of 1.6\% versus 2.6\%
for the minimum second moment criterion. It should be emphasized
that the computational cost of the new algorithm is orders of
magnitude slower than that of a full simulation. To our surprise
the final packing can be computed quite accurately without ever
knowing the shape of the liquid surface during the packing
process!

Finally, we remark on the minimal moment criterion itself: our
results suggest that the drying influences the final packings only
through (a) enforcing the initial disk packing at the critical
volume; and (b) through equation (\ref{fd}), relating the $F_i$ to
the particle displacements. It is probably not coincidental that
(up to prefactors) equation (\ref{fd}) is similar to the
force-displacement relation $F_i = -\partial_i {\cal M} \sim c_1
\delta r_i + c_2 \sum_i \delta r_i$, for the minimal moment
criterion. We believe that this explains the similarity between
the observed structures and those minimizing the second moment.

The calculations presented herein suggest that the unique packings
observed by Manoharan {\it et. al.} arise because (i) the initial
circular packing is unique for the regime they explored, except
for $N=15$; and (ii) the subsequent evolution of the particles is
so highly constrained that there is only one final packing that is
consistent with the constraints.

\begin{figure}[t]
\includegraphics[width=.48\textwidth]{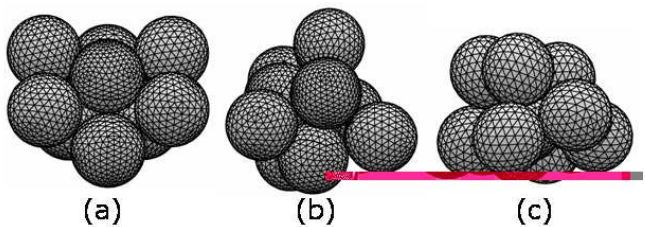}
\centering \caption{Final configuration of 9 spheres with
different wetting conditions: on the droplet, $n_1$ spheres have a
contact angle of 160\deg\, and $n_2=9-n_1$ spheres have a contact
angle of 20\deg. (a) cases $\{n_1,n_2\}=\{2,7\}$ and $\{4,5\}$,
${\cal M}= 27.706 $; (b) case $\{n_1,n_2\}=\{6,3\}$, ${\cal M}=
29.780$; (c) case $\{n_1,n_2\}=\{8,1\}$, ${\cal M}= 30.754$ (units
of the particle radius).} \label{differentwetting}
\end{figure}

This suggests that the only way to generate different packings is
to modify the circle packing at the critical volume. This can be
easily modified by choosing particles with differing sizes or
wettabilities (thus creating circles of different sizes). We have
run simulations where particles on a given droplet possess
different contact angles. Figure \ref{differentwetting} shows
three cases of $N=9$ particles with contact angles of either
160\deg or 20\deg. The three packings and their second moment
differ significantly from Fig. \ref{model}.

In summary, we have presented a numerical and theoretical study of
the packing selection problem of Manoharan {\it et al.}
\cite{Vinny}. The selection of a unique packing was found to arise
almost entirely from geometrical constraints during the drying
process and a procedure was proposed to generate different
packings. This could be extended to characterize all admissible
evaporation-driven packings of colloidal spheres.

We thank  Henry Chen, Vinothan Manoharan,  David Pine and Howard
Stone for useful discussions. We also thank Ken Brakke for his
help with The Surface Evolver. We gratefully acknowledge the
Office of Naval Research and the Harvard MRSEC for supporting this
research.


\begin{thebibliography}{}

\bibitem[Manoharan {\it et al.} (2003)]{Vinny}
{V.N. Manoharan, M.T. Elsesser, and D.J. Pine } (2003) {\it
Science} {\bf 301}, 483-487.

\bibitem[Sloane {\it et al.} (2003)]{Sloane}
Sloane N.J.A., Hardin R.H., Duff T.S. \& Conway J.H. (1995) {\it
Disc. Comp. Geom.} {\bf 14} 237-259. Packings available at
www.research.att.com/~njas/packings/

\bibitem[Kralchevsky P.A. \& Denkov N.D. (2001)]{review}
{Kralchevsky P.A. \& Denkov N.D.} (2001) {\it Curr. Opin. Colloid.
Int. Sci.} {\bf 6} 383-401.

\bibitem[Brakke (1992)]{Brakke}
Brakke K. (1992) {\it Exp Math.} {\bf 1} 141-165.

\bibitem[Melnyk et al (1977)]{Melnyk}
Melnyk T.W., Knop O. \& Smith W.R. (1977) {\it Can. J. Chem. }
{\bf 55} 1745-1761.

\bibitem[Clare (1986)]{Clare}
Clare B.W. \& Kepert D.L. (1986) {\it Proc. Roy. Soc.} A {\bf 405}
329-344.

\bibitem[Sloane et al (2004)]{Sloane_circles}
Sloane N.J.A. with the collaboration of Hardin R.H., Smith W.D.
and others, {\it Tables of Spherical Codes}, published
electronically at www.research.att.com/~njas/packings/

\bibitem[Kottwitz (1991)]{Kottwitz}
Kottwitz D.A. (1991) {\it Acta. Cryst.} A {\bf 47} 158-165.

\bibitem[Witten  et al (1999)]{Witten}
{Tkachenko A.V.  and Witten T.A.} (1999)  {\it Phys. Rev. E}
{\bf 60} 687-696.

\bibitem[Pieranski (1980)]{Pieranski}
{Pieranski P.} (1980) {\it Phys. Rev. Lett.} {\bf 45} 569-572.

\bibitem[Hurd (1985)]{Hurd}
{Hurd A.J.} (1985) {\it J. Phys. A: Math. Gen.} {\bf 18}
L1055-1060.

\end{thebibliography}
\end{document}